\documentclass[sigconf]{acmart}

\usepackage{colortbl}
\usepackage[capitalise]{cleveref}
\usepackage{svg}
\usepackage{makecell}
\usepackage{subcaption}
\usepackage{graphicx}

\newcommand{\gcotwoe}{\emph{gCO\textsubscript{2}e} }

\AtBeginDocument{%
  \providecommand\BibTeX{{%
    \normalfont B\kern-0.5em{\scshape i\kern-0.25em b}\kern-0.8em\TeX}}}

\copyrightyear{2024}
\acmYear{2024}
\setcopyright{rightsretained}
\acmConference[RecSys '24]{18th ACM Conference on Recommender Systems}{October 14--18, 2024}{Bari, Italy}
\acmBooktitle{18th ACM Conference on Recommender Systems (RecSys '24), October 14--18, 2024, Bari, Italy}\acmDOI{10.1145/3640457.3688074}
\acmISBN{979-8-4007-0505-2/24/10}

\begin{document}

\title{From Clicks to Carbon: \\The Environmental Toll of Recommender Systems}

\author{Tobias Vente}
\authornote{Both authors contributed equally to this research.}
\email{tobias.vente@uni-siegen.de}
\orcid{0009-0003-8881-2379}
\affiliation{%
  \institution{Intelligent Systems Group, University of Siegen}
  \streetaddress{Adolf-Reichwein-Straße 2a}
  \city{Siegen}
  \country{Germany}
  \postcode{57076}
}
\author{Lukas Wegmeth}
\authornotemark[1]
\email{lukas.wegmeth@uni-siegen.de}
\orcid{0000-0001-8848-9434}
\affiliation{%
  \institution{Intelligent Systems Group, University of Siegen}
  \streetaddress{Adolf-Reichwein-Straße 2a}
  \city{Siegen}
  \country{Germany}
  \postcode{57076}
}
\author{Alan Said}
\email{alan@gu.se}
\orcid{0000-0002-2929-0529}
\affiliation{%
  \institution{University of Gothenburg}
  \city{Gothenburg}
  \country{Sweden}
}
\author{Joeran Beel}
\email{joeran.beel@uni-siegen.de}
\orcid{0000-0002-4537-5573}
\affiliation{%
  \institution{Intelligent Systems Group, University of Siegen}
  \streetaddress{Adolf-Reichwein-Straße 2a}
  \city{Siegen}
  \country{Germany}
  \postcode{57076}
}

\renewcommand{\shortauthors}{Vente and Wegmeth, et al.}



\begin{abstract}
As global warming soars, the need to assess the environmental impact of research is becoming increasingly urgent. 
Despite this, few recommender systems research papers address their environmental impact. 
In this study, we estimate the environmental impact of recommender systems research by reproducing typical experimental pipelines. 
Our analysis spans 79 full papers from the 2013 and 2023 ACM RecSys conferences, comparing traditional ``good old-fashioned AI’' algorithms with modern deep learning algorithms.
We designed and reproduced representative experimental pipelines for both years, measuring energy consumption with a hardware energy meter and converting it to CO\textsubscript{2} equivalents. 
Our results show that papers using deep learning algorithms emit approximately 42 times more CO\textsubscript{2} equivalents than papers using traditional methods. 
On average, a single deep learning-based paper generates 3,297 kilograms of CO\textsubscript{2} equivalents---more than the carbon emissions of one person flying from New York City to Melbourne or the amount of CO\textsubscript{2} one tree sequesters over 300 years.

\end{abstract}

\maketitle

\section{Introduction}
Carbon emissions play a pivotal role in global warming by driving the greenhouse effect that leads to rising temperatures and extreme weather events \cite{nukusheva2021global, kweku2018greenhouse, mitchell1989greenhouse, hansen2013assessing, sinha2019review}. 
With the ambitious goal of the \emph{United Nations Framework Convention on Climate Change} to cap the global warming temperature increase at 1.5 degrees Celsius\footnote{\url{https://unfccc.int/documents/184656}}, reducing carbon emissions becomes crucial \cite{hansen2013assessing}.
Therefore, significant reductions in carbon emissions are urgently needed to meet the temperature target and mitigate the climate impact \cite{hansen2013assessing, sinha2019review, yoro2020co2}.

Concurrently, computationally heavy algorithms have become the norm for modern recommender systems, increasing the energy consumption of recommender systems experiments \cite{Chen_Zhang_Zhang_Dai_Yi_Zhang_Zhang_2020}. 
This trend is driven by a shift from traditional algorithms like \emph{ItemKNN} (or so-called \emph{good old-fashioned AI}) to more sophisticated deep learning techniques \cite{10.1145/3285029, 8529185}.
The increased energy consumption of deep learning results in higher carbon emissions, further exacerbating environmental challenges.
With this backdrop, a relevant question is \emph{what is the environmental toll of recommender systems experiments?}

However, the recommender systems community has paid little attention to the energy consumption and carbon emissions of their experiments \cite{10.1145/3604915.3608840}.
This oversight is increasingly concerning, given the global consensus on the urgent need for actions to mitigate carbon emissions \cite{1130574323982454028}.
As the recommender systems community strives to advance technologies that improve user experiences, it must also confront the environmental repercussions and their impact on climate change, especially during the current crisis.

\phantomsection
\label{para:researchQuestions} 
Therefore, our interest is to answer the question: What are the ecological costs of recommender systems research, past and present?
In this paper, we answer the following research questions.

\begin{enumerate}
    \item[\textbf{RQ1:}] How large is the energy consumption of a modern recommender systems research paper?
    \item[\textbf{RQ2:}] How substantial is the energy consumption and performance trade-off between traditional and deep learning recommender system algorithms?
    \item[\textbf{RQ3:}] How has the carbon footprint of recommender systems experiments changed with the transition from traditional to deep learning algorithms?
\end{enumerate}

In this work, we reproduce \emph{representative} recommender systems experimental pipelines and show the extent of carbon emissions attributable to recommender systems experiments, providing a comparative analysis of recommender systems algorithms a decade apart.
Our analysis is based on 79 full papers from the ACM RecSys conference in 2013 and 2023, respectively. 
Reproducing a representative experiment setup for each year enables us to conduct these experiments to directly measure their energy consumption and determine their carbon footprint across various hardware configurations, including laptops, workstations, and desktop PCs. 

Our contribution is a comprehensive analysis of the carbon emissions associated with recommender system experiments and the reproduction of a representative recommender systems pipeline comparing 13 datasets and 23 algorithms from 2013 and 2023 on four hardware configurations. 
Our results indicate that a recommender systems research paper utilizing deep learning algorithms produces, on average, 3,297 kilograms of CO\textsubscript{2} equivalents.
In contrast to a paper utilizing traditional algorithms from 2013, a 2023 research paper emits approximately 42 times more CO\textsubscript{2} equivalents.
Furthermore, we show that the geographical location, i.e., the means of energy production (fossil or renewable), can change the carbon footprint of recommender systems research experiments by up to 12 times. 
Additionally, we found vast differences in the energy efficiency of different hardware configurations when performing recommender systems experiments, changing the carbon footprint of the same experiments by a factor of up to 10.
We highlight the energy consumption and carbon emissions of recommender system experiments to foster awareness and the development of more sustainable research practices in the field, ultimately contributing to more sustainable recommender systems experiments and research. 

The code used to execute and measure the experiments is publicly available in our GitHub\footnote{\url{https://github.com/ISG-Siegen/recsys-carbon-footprint}} repository and further contains documentation to ensure the reproducibility of our experiments.
\section{Related Work}

The impacts of greenhouse gases on the environment are extensively studied, with measurable effects seen in global warming and climate change. 
This research underscores the importance of addressing greenhouse gases for the well-being of humanity and life on Earth.
According to meta-studies, scientists overwhelmingly agree that humans cause climate change \cite{Lynas2021, Cook2016}.
Additionally, tremendous efforts are being made to document the development of greenhouse gas emissions globally\footnote{\url{https://edgar.jrc.ec.europa.eu/report_2023}}.
The greenhouse gas carbon dioxide (CO\textsubscript{2})~\cite{Cain2019} is emitted in the generation of electricity when using fossil fuels\footnote{\url{https://www.iea.org/reports/co2-emissions-in-2022}}.
Particularly relevant to applied researchers, optimistic estimations forecast that more than 7\% of global electricity demand will be attributed to computing by 2030 \cite{9407142}.

Electricity carries a direct monetary value and a cost associated with the damage caused by greenhouse gas emissions from its production, commonly referred to as \emph{social cost} \cite{Ricke2018}.
Hence, due to the harmful effects of fossil fuel electricity production, there is growing interest in green computing \cite{10214579}.
In machine learning literature, energy consumption and carbon emissions from research experiments are well-known concerns \cite{budennyy2022eco2ai, mehta2023review}.
Green computing in machine learning presents unique challenges \cite{van2021sustainable}, particularly due to the shift to deep learning techniques, including, e.g., natural language processing \cite{strubell2019energy} and computer vision  \cite{https://doi.org/10.48550/arxiv.2311.00447}.
Natural language processing tasks are frequently solved with large language models in cutting-edge recommender systems research \cite{wu2023survey}.

The machine learning community actively provides guidelines and raises awareness for green and sustainable computing \cite{Lannelongue2023, JMLR:v21:20-312}.
Numerous software libraries have also been published to measure the carbon footprint of existing experiments \cite{powermeter}.
However, despite these advancements, green and sustainable computing have yet to become established topics within recommender systems.

To our knowledge, only one paper has directly examined the carbon footprint of recommender systems \cite{10.1145/3604915.3608840}.
This study, however, is limited to exploring the trade-off between algorithm performance and carbon footprint, relying on software-based power measurements rather than hardware-based analysis.
Research on automated recommender systems has considered computing power requirements \cite{Wegmeth2022-bx,vente2023introducing}, significantly impacting energy consumption.
Additionally, another paper proposes an energy-efficient alternative to k-fold cross-validation \cite{beel_wegmeth_vente_2024}, and recommender systems have been explored for their potential to enhance sustainability and energy efficiency in other domains \cite{HIMEUR20211,af69af4ea12246b6b0c0d73908ed1fc5}.
However, accurately estimating the global carbon footprint of recommender systems remains impossible without further research.

Our literature review reveals that current recommender systems papers do not openly disclose the estimated carbon footprint of their experiments.
Furthermore, to our knowledge, no recommender systems publication outlets publicly release statements regarding the carbon footprint of submissions.
An example of this in this area is ECIR 2024, which incorporated self-reported greenhouse gas emissions into its paper submissions, though no official results have been published yet.
We conclude that while awareness of the carbon footprint of recommender systems experiments is increasing, it remains insufficiently low to make a significant impact.

\section{Comparative Study: Recommender Systems Research in 2023 versus 2013}\label{paper_review}
To answer our research questions, we reproduce measurements and estimations as accurately as possible.
Therefore, we summarize the historical development of recommender systems experiments by analyzing research papers from 2023 and 2013, reflected through peer-reviewed papers at a conference.
To this end, we present our analysis of all \emph{full papers} accepted in the main track at ACM RecSys in 2013 (32 papers~\cite{10.1145/2507157}) and 2023 (47 papers~\cite{10.1145/3604915}).
All papers considered in this analysis are published in the ACM Digital Library.

In the following paragraphs, we examine these recommender systems papers for \textbf{hardware} specifications, software \textbf{libraries}, design decisions in \textbf{experimental pipelines}, the availability of \textbf{open-source code} for reproducibility, and used \textbf{datasets}.

\paragraph{Hardware:} 
Only 15 (32\%) full papers from ACM RecSys 2023 detail the hardware used for experiments, and consequently, 32 (68\%) of the papers do not report this information.
All these 15 papers explicitly report usage of Nvidia GPUs, with seven specifically mentioning the Nvidia v100 GPU (released in 2017). 

Looking back at papers published at ACM RecSys 2013, only 6 out of 32 (19\%) contain information about the hardware used, meaning 26 (81\%) do not contain hardware information.
Contrary to 2023, none of the papers from 2013 mention using a GPU. 
However, all 6 papers that disclose their hardware utilize an Intel Xeon CPU.

\paragraph{Libraries:} 
At ACM RecSys 2023, 18 (38\%) papers use PyTorch, and 6 (13\%) papers use TensorFlow implementing deep learning recommender systems. 
RecBole \cite{recbole[1.2.0]} is used by 5 (11\%) papers, making it the most popular library designed explicitly for recommender systems. 
While 13 (28\%) papers from 2023 do not specify the library used, they all implement deep learning algorithms.
The remaining 5 (11\%) papers report using one of the following libraries: Elliot \cite{DBLP:conf/sigir/AnelliBFMMPDN21}, ReChorus \cite{wang2020make}, Bambi \cite{Capretto2022}, FuxiCTR \cite{DBLP:conf/sigir/ZhuDSMLCXZ22}, or CSRLab \cite{crslab}.

This contrasts the patterns observed in ACM RecSys 2013 papers. 
The majority of papers, 27 (84\%) do not report using any open libraries. 
Instead, they rely on private algorithm implementations.
Only 5 (16\%) papers report using libraries, which include MyMediaLite \cite{Gantner2011MyMediaLite}, Apache Mahout \cite{10.5555/3455716.3455843}, InferNet \cite{InferNET18}, and libpMF \cite{hfy12a}. 
This shift highlights a considerable evolution in adopting standardized libraries within recommender systems over the past decade. 
Arguably, the feasibility of reproducing a recommender systems paper has increased over time, at least in terms of software.

\paragraph{Experimental Pipeline:} 
Between papers from 2023 and 2013, the experimental pipeline remains consistent, but there are significant differences in the algorithms and datasets.
For instance, the holdout split is the most popular data splitting technique, used in 20 (42\%) of the 2023 papers and 20 (63\%) of the 2013 papers. 
Grid search is also the favored optimization technique in both 2023 and 2013.
Additionally, 22 (47\%) papers in 2023 and 20 (63\%) in 2013 do not employ dataset pruning, although n-core pruning appears to have gained notable popularity, appearing in 19 (40\%) papers in 2023.

One of the most significant differences lies in the evaluation metrics used: nDCG is the predominant metric in 2023, used in 32 (68\%) papers, whereas, in 2013, Precision is the most popular one, appearing in 13 (40\%) papers, followed by RMSE and nDCG, each used in 5 (16\%) papers.
This is partly because 44 (94\%) of papers in 2023 focus on top-n ranking prediction tasks, while 18 (56\%) of papers in 2013 focus on rating prediction tasks.

\paragraph{Open-Source Code:}
Only 18 (38\%) of ACM RecSys 2023 papers do not contain links to their source code. 
On the other hand, 29 (62\%) make their code available, with all but 2 hosting it on \emph{GitHub} and the remaining hosting it on their organization's website.
However, 3 (6\%) repositories linked in these papers are empty or unreachable.
Only 1 (3\%) paper from ACM RecSys 2013 shares code. 

\paragraph{Datasets:} 
On average, papers from ACM RecSys 2023 include three datasets. 
The most frequent datasets are from the Amazon2018 series, appearing in 15 (32\%) papers, followed by the MovieLens datasets, of which MovieLens-1M is used in 11 (23\%) papers, while MovieLens-100K, MovieLens-10M, and MovieLens-20M are used in 2 (4\%) papers. 
Other commonly used datasets are LastFM in 6 (13\%) papers, Yelp-2018 in 5 (11\%) papers, and Gowalla in 5 (11\%) papers.

On average, papers at ACM RecSys 2013 include two distinct datasets in their experimental pipeline. 
The most frequently used datasets are the MovieLens datasets, used in 10 (31\%) papers. 
Neither of the popular Amazon datasets were available in 2013. 
Other commonly used datasets include the LastFM dataset, used in 4 (13\%) papers, and the Netflix Prize dataset, used in 3 (9\%) papers. 
\section{Method}
We measure the energy consumption of running 23 algorithms on 13 datasets with a smart power plug.
To assess the impact of hardware on energy efficiency, we run the experiments on four distinct computers.
We also estimate the carbon footprint by assessing the carbon emissions related to electricity generation in five locations. 
Our experimental pipeline is based on our research papers analysis, ensuring representative data collection (\cref{paper_review}). All design decisions are derived from this analysis unless otherwise specified.

\subsection{Experimental Pipeline}
We randomly divide each of the 13 datasets into three splits, where 60\% of the data is for training, 20\% for validation, and 20\% for testing. 
Since our work focuses on energy consumption rather than maximizing performance and generalizability, we neither employ cross-validation nor repeat experiments. 
While this decision comes at the cost of reliability and performance in terms of accuracy, our goal is not to optimize the recommender models to beat a baseline but to measure the power consumption of a characteristic recommender system experiment. 
We measure performance by nDCG@10 for top-n ranking predictions and RMSE for rating predictions.
Furthermore, we use the default hyperparameter settings provided by the libraries instead of optimizing them.
We make these decisions to minimize unnecessary energy consumption.
All deep learning algorithms are trained for 200 epochs, with model validation after every fifth epoch to facilitate early stopping.

\subsubsection{Datasets:}
Based on our research paper analysis (\cref{paper_review}), we include 13 datasets in our experiments and refer to \cref{tab:dataset_statistics} for an overview.
For top-n prediction tasks, we convert rating prediction datasets according to practice that is common in our paper analysis \cite{10.1145/3604915.3608802,10.1145/3604915.3608791,10.1145/3604915.3608771,10.1145/3604915.3608785}.
Furthermore, we prune all datasets such that all included users and items have at least five interactions, commonly known as five-core pruning \cite{10.1145/3357384.3357895,10.1145/3460231.3474275,10.1145/3523227.3546770}.
\cref{tab:dataset_statistics} shows the dataset statistics of all included datasets for the preprocessed datasets.

\renewcommand{\arraystretch}{1.2}
\begin{table}
\caption{Basic information of the data sets used in our experiments after preprocessing.}
\resizebox{\linewidth}{!}{
\begin{tabular}{l|l|l|l|l}
\toprule
Dataset Name & Users & Items & Interactions & Density \\
\midrule
\makecell[tl]{Amazon2018\cite{ni-etal-2019-justifying}\\Books} & 105,436 & 151,802 & 1,724,703 & 0.0108 \\
\makecell[tl]{Amazon2018\cite{ni-etal-2019-justifying}\\CDs-And-Vinyl} & 71,943 & 107,546 & 1,377,008 & 0.0178 \\
\makecell[tl]{Amazon2018\cite{ni-etal-2019-justifying}\\Electronics} & 62,617 & 187,288 & 1,476,535 & 0.0126 \\
\makecell[tl]{Amazon2018\cite{ni-etal-2019-justifying}\\Sports-And-Outdoors} & 69,781 & 185,024 & 1,498,609 & 0.0116 \\
\makecell[tl]{Amazon2018\cite{ni-etal-2019-justifying}\\Toys-And-Games} & 75,856 & 192,326 & 1,686,250 & 0.0116 \\
Gowalla\cite{10.1145/2020408.2020579} & 64,115 & 164,532 & 2,018,421 & 0.0191 \\
Hetrec-LastFM\cite{Cantador:RecSys2011} & 1,090 & 3,646 & 52,551 & 1.3223 \\
\makecell[tl]{MovieLens\cite{10.1145/2827872}\\100K} & 943 & 1,349 & 99,287 & 7.8049 \\
\makecell[tl]{MovieLens\cite{10.1145/2827872}\\1M} & 6,040 & 3,416 & 999,611 & 4.8448 \\
\makecell[tl]{MovieLens\cite{10.1145/2827872}\\Latest-Small} & 610 & 3,650 & 90,274 & 4.0545 \\
Netflix\footnote{\url{https://www.kaggle.com/datasets/netflix-inc/netflix-prize-data}} & 11,927 & 11,934 & 5,850,559 & 4.1103 \\
Retailrocket\footnote{\url{https://www.kaggle.com/datasets/retailrocket/ecommerce-dataset}} & 22,178 & 17,803 & 240,938 & 0.0610 \\
Yelp-2018\footnote{\url{https://www.yelp.com/dataset}} & 213,170 & 94,304 & 3,277,931 & 0.0163 \\
\bottomrule
\end{tabular}
}
\label{tab:dataset_statistics}
\end{table}
\renewcommand{\arraystretch}{1.0}

\subsubsection{Algorithms:}
Based on our research paper analysis (\cref{paper_review}), we include 23 frequently used algorithms in our experiments. 
The algorithm implementations are from RecBole \cite{recbole[1.2.0]} (indicated by \emph{RB}), RecPack \cite{10.1145/3523227.3551472}  (indicated by \emph{RP}) and LensKit \cite{ekstrand2020lenskit} (indicated by \emph{LK}).
We run algorithms from RecBole on a GPU while we run algorithms from LensKit and RecPack on a CPU. 
\cref{tab:algorithms} shows all algorithm implementations used in our experiments.

\begin{table}
\caption{Algorithm implementations used in our experiments.}
\resizebox{\linewidth}{!}{
\begin{tabular}{l|l|l}
\toprule
Library & Executed on & Algorithm \\
\midrule
RecBole & GPU & \makecell[tl]{BPR, DGCF, DMF, ItemKNN, LightGCN,\\ MacridVAE, MultiVAE, NAIS, NCL, NeuMF,\\ NGCF, Popularity, RecVAE, SGL}  \\
\midrule
RecPack & CPU & \makecell[tl]{ItemKNN, NMF, SVD} \\
\midrule
LensKit & CPU & \makecell[tl]{ImplicitMF, ItemKNN, UserKNN, Popularity,\\ BiasedMF, FunkSVD} \\
\bottomrule
\end{tabular}
}
\label{tab:algorithms}
\end{table}

\subsection{Representative Pipelines}
Our research paper study indicates that experimental pipelines from 2013 and 2023 exhibit notable differences (\cref{paper_review}). 
For instance, in 2023, all but one paper focused on top-n ranking prediction tasks, whereas in 2013, around half of the papers focused on rating prediction tasks. 
Additionally, in 2023, experiments often utilize datasets from the \emph{Amazon-2018} series, which were not available in 2013. 
Consequently, we introduce three distinct representative pipelines to account for these differences.
\cref{tab:specific_pipelines} provides an overview of the algorithms and datasets used for specific pipelines. 

\begin{table}
\caption{Representative pipelines used to run experiments.}  
\resizebox{\linewidth}{!}{
\begin{tabular}{l|l|l}
\toprule
\makecell[tl]{Year and\\ Prediction Type} & Algorithms & Datasets \\
\midrule
\makecell[tl]{2013\\Rating\\Prediction} & \makecell[tl]{ItemKNN$^{LK}$, UserKNN$^{LK}$,\\ BiasedMF$^{LK}$, FunkSVD$^{LK}$} & \makecell[tl]{Movielens-100K,\\ Movielens-1M,\\ Netflix} \\
\midrule
\makecell[tl]{2013\\Top-N\\Ranking\\Prediction} & \makecell[tl]{ImplicitMF$^{LK}$, ItemKNN$^{LK}$,\\ UserKNN$^{LK}$, Popularity$^{LK}$,\\ ItemKNN$^{RP}$, NMF$^{RP}$, SVD$^{RP}$,\\ BPR$^{RB}$, ItemKNN$^{RB}$,\\ Popularity$^{RB}$} & \makecell[tl]{Hetrec-Lastfm,\\ Movielens-100K,\\ Movielens-1M,\\ Gowalla} \\  
\midrule
\makecell[tl]{2023\\Top-N\\Ranking\\Prediction} & 
\makecell[tl]{ImplicitMF$^{LK}$, ItemKNN$^{LK}$,\\ UserKNN$^{LK}$, Popularity$^{LK}$,\\ ItemKNN$^{RP}$, NMF$^{RP}$, SVD$^{RP}$,\\ BPR$^{RB}$, DGCF$^{RB}$, DMF$^{RB}$,\\ ItemKNN$^{RB}$, LightGCN$^{RB}$,\\ MacridVAE$^{RB}$, MultiVAE$^{RB}$,\\ NAIS$^{RB}$, NCL$^{RB}$, NeuMF$^{RB}$,\\ NGCF$^{RB}$, Popularity$^{RB}$,\\ RecVAE$^{RB}$, SGL$^{RB}$} & \makecell[tl]{Gowalla,\\ Hetrec-LastFM,\\ MovieLens:\\-100K,\\-1M,\\-Latest-Small,\\ Amazon2018:\\-Electronics,\\-Toys-And-Games,\\-CDs-And-Vinyl,\\-Sports-And-Outdoors,\\-Books,\\ Yelp-2018,\\ Retailrocket}\\
\bottomrule
\end{tabular}
}
\label{tab:specific_pipelines}
\end{table}

\subsection{Calculating Greenhouse Gas Emission}
To calculate the greenhouse gas emissions from recommender systems experiments, we first record the electricity consumption and translate the consumption into equivalent emissions. 

\subsubsection{Measuring Electrical Energy Consumption}
To precisely measure the energy consumption of recommender system experiments, we equip each computer in our experimental setup with a commercially available off-the-shelf smart power plug\footnote{\url{https://www.shelly.com/en-se/products/product-overview/shelly-plus-plug-s}}.
This enables us to measure energy consumption in kilowatt-hours (\emph{kWh}) at half-second intervals.
We then align the power plug's measurements with the experiments based on timestamps.
Our measurements capture the entire hardware's energy usage, covering components such as cooling, power supply, CPU, GPU, and memory.

\subsubsection{Calculating Greenhouse Gas Emissions Based on Electricity Consumption}
We convert the measured energy consumption in \emph{kWh} into carbon dioxide equivalents (\emph{CO\textsubscript{2}e}) utilizing the comprehensive dataset provided by \emph{Ember}\footnote{\url{https://ember-climate.org/}}. 
The Ember datasets merge data from the European Electricity Review and information from various original data providers into a singular dataset\footnote{\url{https://docs.owid.io/projects/etl/}}. 

The Ember dataset features a conversion rate from \emph{kWh} to grams of carbon dioxide emissions equivalents (\emph{gCO\textsubscript{2}e}). 
\emph{CO\textsubscript{2}e} represent the greenhouse gas emissions released and the overall environmental impact of electricity generation. 
Since the hardware does not directly impact the environment by, e.g., emitting greenhouse gases, we utilize the carbon dioxide equivalents associated with the electricity generation process of the consumed energy.

The \emph{gCO\textsubscript{2}e} conversion rate linked to electricity generation varies notably based on the method of production \cite{kerem2022investigation}.
Electricity sourced from renewable energy, such as hydropower, typically has a lower carbon footprint than coal combustion.
Different regions employ diverse energy generation methods, so we use the global average conversion rate and compare various geographical locations.

\subsection{Hardware}
We conduct all experiments across four computers with different hardware configurations from different years to assess the impact of hardware efficiency on energy consumption and, consequently, the carbon footprint.
Computer hardware has become increasingly efficient over the years \cite{5440129}. 
We perform experiments on four computers spanning the last decade to evaluate the impact on hardware efficiency and whether the improvement can offset the energy demands of transitioning to deep learning. 
We present the hardware specifications used in our experiments in \cref{tab:hardware}.

\begin{table}
\caption{Hardware used to compare hardware efficiency.}  
\resizebox{\linewidth}{!}{
\begin{tabular}{l|l|l|l|l}
\toprule
\makecell[tl]{Computer and\\Year} & CPU & GPU & \makecell[tl]{RAM\\in GB} & \makecell[tl]{Storage\\in TB} \\
\midrule
\makecell[tl]{Modern Workstation \\ 2023} & \makecell[tl]{Intel Xeon\\W-2255\\@ 3.70 GHz} & \makecell[tl]{NVIDIA\\GeForce\\RTX 3090} & 256 & 2 \\
\midrule
\makecell[tl]{Mac Studio \\2022} & M1 Ultra & M1 Ultra & 64 & 1 \\
\midrule
\makecell[tl]{MacBook Pro \\2020} & M1 & M1 & 16 & 1 \\
\midrule
\makecell[tl]{Legacy Workstation \\2013} & \makecell[tl]{Intel Core\\i7-6700K \\@ 4.00GHz} & \makecell[tl]{NVIDIA\\GeForce\\GTX 980 Ti} & 128 & 1 \\
\bottomrule
\end{tabular}
}
\label{tab:hardware}
\end{table}
\section{Results}
Our results demonstrate the energy consumption, the trade-offs between energy and performance, and the carbon footprint associated with the 2013 and 2023 ACM RecSys full paper experiments.

\subsection{The energy consumption of a 2023 recommender systems research paper} \label{RQ1}

\begin{figure*}
    \centering
    \begin{subfigure}[b]{0.45\textwidth}
         \includegraphics[width=1\linewidth]{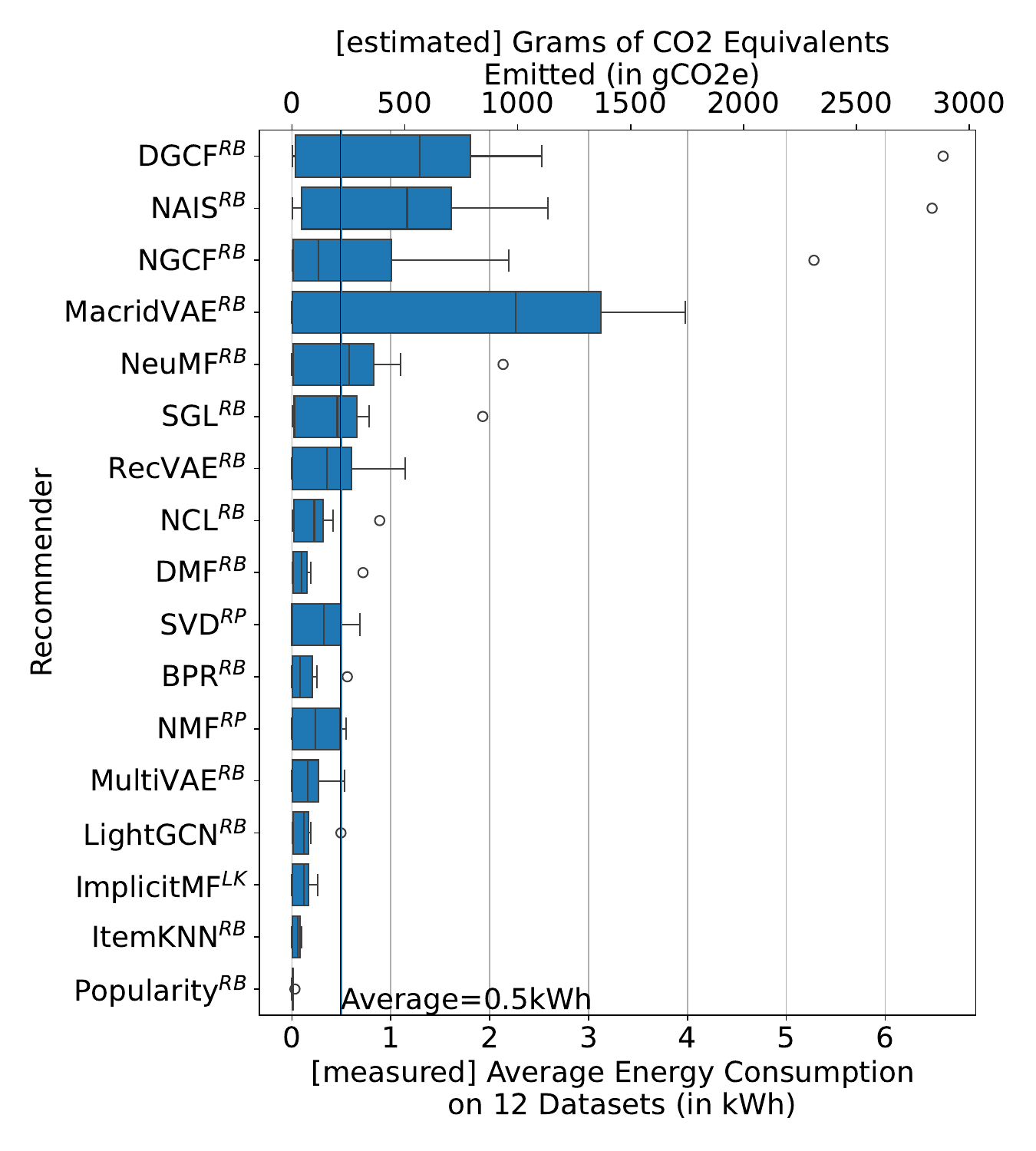}
         \caption{Energy Consumption for 16 Algorithms}
         \label{fig:avgalgo}
    \end{subfigure}\hfill
    \begin{subfigure}[b]{0.55\textwidth}
        \includegraphics[width=1\linewidth]{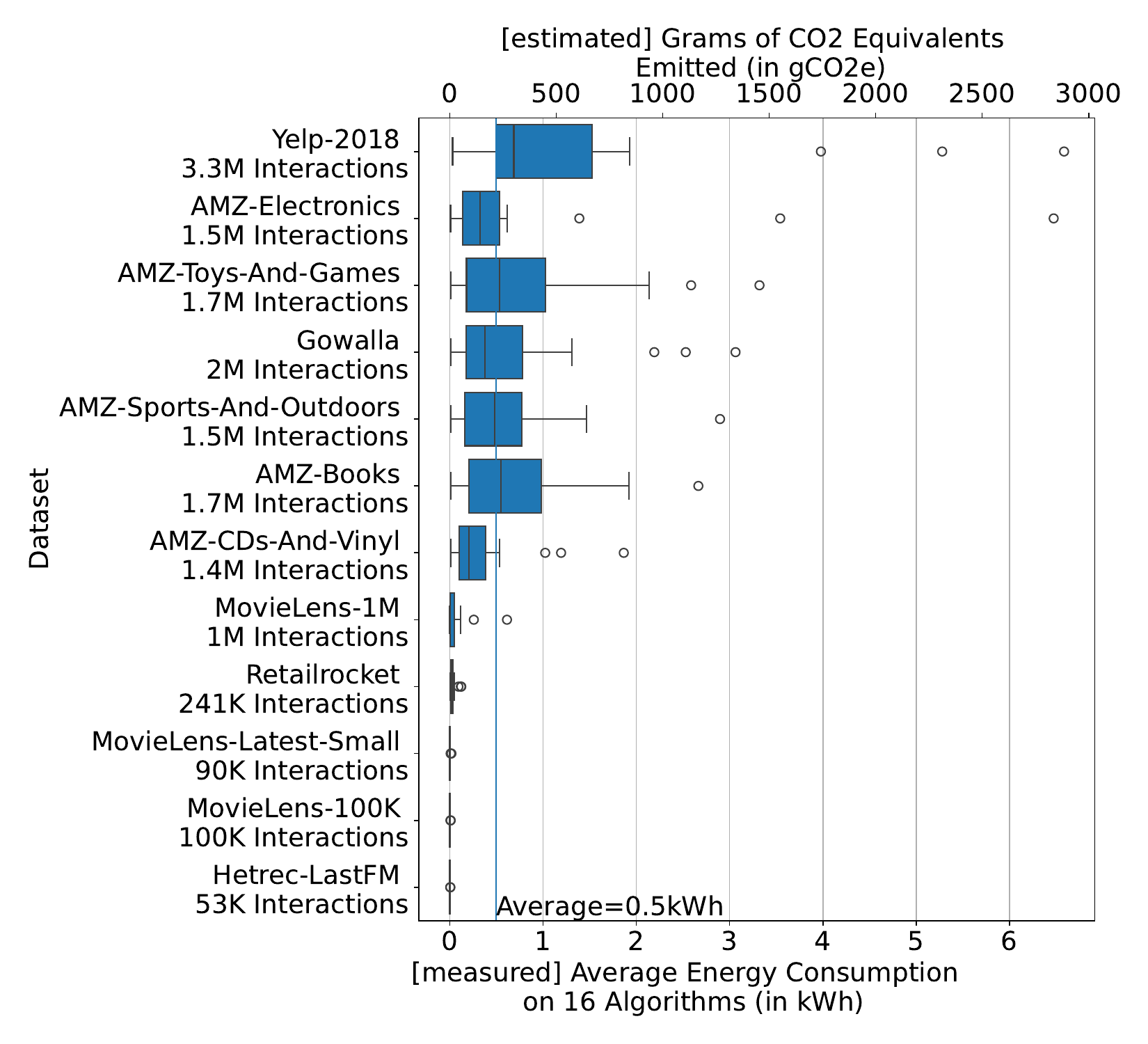}
        \caption{Energy Consumption for 12 Datasets}
        \label{fig:avgdata}
        
    \end{subfigure}\hfill
    \caption{Average power consumption of recommender system algorithms across twelve datasets, alongside the average consumption per dataset across sixteen algorithms. \emph{Blue} vertical lines represent the average consumption in \emph{kWh}. The upper x-axis displays the equivalent CO\textsubscript{2} emissions in grams (\emph{gCO\textsubscript{2}e}), calculated based on the 2023 world average.}
    \Description[Test]{Test}
    \label{fig:avgpowcons}
\end{figure*}

\subsubsection{The Energy Consumption of an Algorithm and Dataset}

Based on our experiments, we estimate the energy consumption of a single run of one recommender systems algorithm on one dataset to be, on average, 0.51 \emph{kWh} (\cref{fig:avgalgo}). 

Energy consumption varies among algorithms and datasets. 
Relatively simple algorithms like \emph{Popularity} and \emph{ItemKNN} consume only 0.007 \emph{kWh} and 0.04 \emph{kWh}, respectively (average over twelve datasets used for experiments representing 2023). 
Some recent deep learning algorithms consume a relatively low amount of energy, e.g., \emph{DMF} and \emph{LightCGN} consume 0.13 \emph{kWh} and 0.12 \emph{kWh}, respectively (average over twelve datasets used for 2023 experiments). 
Contrasting, the most ``expensive'' algorithms -- \emph{MacridVAE} and \emph{DCGF} consume, on average, 1.79 \emph{kWh} and 1.45 \emph{kWh}, respectively. 
Consequently, the most expensive algorithm (MacridVAE), regarding electricity consumption, requires 257 times as much energy as the cheapest algorithm (Popularity). 

The energy consumption of individual algorithms on different datasets is even higher.
\emph{Popularity} consumes only 0.000036 \emph{kWh} on \emph{Movielens-Latest-Small} but 0.03 \emph{kWh} on the \emph{Yelp-2018} dataset (factor 800). 
The deep learning algorithm \emph{DGCF} consumes 0.005 \emph{kWh} on \emph{Hetrec-LastFM} but 6.6 \emph{kWh} on \emph{Yelp-2018} (factor 1,444). 

When executing algorithms over larger datasets, energy consumption increases compared to executing the same algorithm on smaller data.
For example, the average recommender systems algorithm consumes 0.001 \emph{kWh} on \emph{Hetrec-LastFM} with 53 thousand interactions and 1.56 \emph{kWh}, 1,560 times more energy, on \emph{Yelp-2018} with 3.3 million interactions (\cref{fig:avgdata}). 

However, the energy consumption per dataset is not solely dependent on the number of interactions. 
Although \emph{Amazon2018-Electronics} has \textasciitilde25\% fewer interactions than \emph{Gowalla} (1.5M vs. 2M; \cref{fig:avgdata}), algorithms running on \emph{Amazon2018-Electronics} consume, on average, 18\% more energy (0.728 \emph{kWh}/0.617 \emph{kWh}). 
Similarly, even though \emph{Movielens-1M} includes around half the interactions of \emph{Gowalla} (1M vs. 2M), algorithms on \emph{Movielens-1M} consume only 10\% of the energy \emph{Gowalla} needs (0.06 \emph{kWh} vs. 0.6 \emph{kWh}). 

\subsubsection{The Energy Consumption of an Experimental Pipeline}\label{ssec:pipeline}
Based on our experiments, we estimate the energy consumption of a representative 2023 recommender systems experimental pipeline to be 171.36 \emph{kWh}.

Through the paper analysis described in \cref{paper_review}, we found that a 2023 recommender systems experimental pipeline includes, on average, seven recommender systems algorithms. 
Additionally, the algorithm performance is, on average, evaluated on three datasets. 
Furthermore, a representative experimental pipeline performs hyperparameter optimization through grid search on 16 configurations per algorithm. 
Since one algorithm consumes, on average, 0.51 \emph{kWh} (\cref{fig:avgpowcons}, left), the energy consumption of an experimental pipeline is calculated as follows:  
 \[ 7 \times 3 \times 16 \times 0.51 \, \emph{kWh} = 171.36 \, \emph{kWh}.\]

\subsubsection{The Energy Consumption of a Paper}
Based on experiments and our paper study (\cref{paper_review}), we estimate the energy consumption of a representative 2023 paper to be 6,854.4 \emph{kWh}.

The energy consumption estimation of 171.36 \emph{kWh} per recommender systems experimental pipeline only accounts for the direct energy consumption during the experimental run (\cref{ssec:pipeline}).
The estimation excludes energy costs for preliminary activities such as algorithm prototyping, initial test runs, data collection, data preprocessing, debugging, and potential re-running of experiments due to pipeline errors.
Therefore, to approximate the total energy impact of a recommender systems paper, we account for these additional energy costs by introducing an additional factor. 

We interviewed the authors of Elliot \cite{DBLP:conf/sigir/AnelliBFMMPDN21}, RecPack \cite{10.1145/3523227.3551472}, LensKit \cite{ekstrand2020lenskit}, and recommender systems practitioners \cite{10.1145/3604915.3609498} asking them to estimate a factor of the energy consumption overhead of a recommender systems paper compared to running the experimental pipeline once.
Their median answer was 40.
We multiplied the energy consumption of experimental pipelines accordingly.

Following this, we estimate the energy consumption of experiments for the results presented in a recommender systems paper ranges to be 6,854.4 \emph{kWh}.

\subsection{Energy Consumption and Performance Trade-Off}

\begin{figure}
    \centering
    \includegraphics[width=1\linewidth]{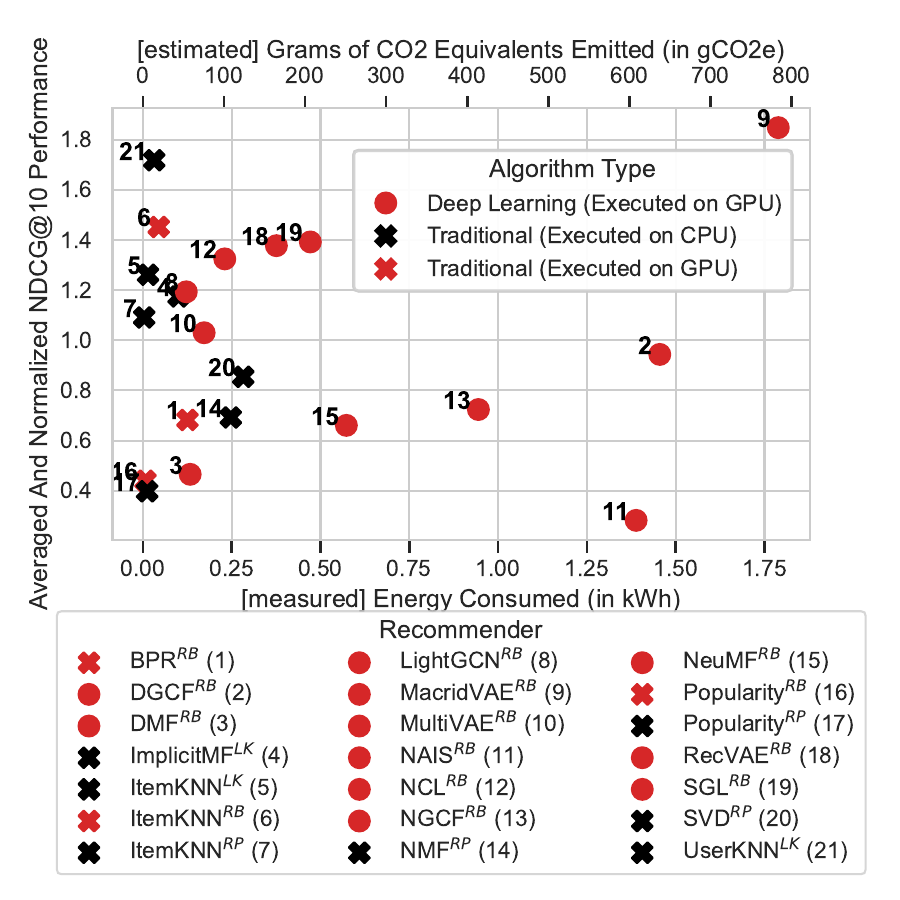}
    \caption{Total energy consumption (in \emph{kWh}) vs. averaged and normalized \emph{nDCG@10} performance. Data points in \textbf{\emph{blue}} represent algorithms running on CPUs and \textbf{\emph{red}} for running on GPUs. A cross represents a traditional, and a dot is a deep learning algorithm. The \emph{nDCG@10} is normalized within each dataset to ensure uniform impact and averaged across all twelve included datasets. The upper x-axis shows the \emph{gCO\textsubscript{2}e} emissions, calculated using the 2023 world average.}
    \Description[Test]{Test}
    \label{fig:relation}
\end{figure}

Our results demonstrate that higher energy consumption does not necessarily lead to better performance (\cref{fig:relation}).
Two of the three top-performing algorithms (\emph{UserKNN$^{LK}$} and \emph{ItemKNN$^{RB}$}) are traditional nearest neighbor algorithms that both consume on average around 0.040 \emph{kWh} for a single run on one dataset (average of twelve datasets used for 2023 experiments). 
In contrast, the third, a deep learning algorithm (\emph{MacridVAE$^{RB}$}) consumes, on average, 45 times more energy at 1.8 \emph{kWh} per run on one dataset.

In general, our results show that deep learning algorithms consume eight times more energy than traditional algorithms for a single run, on average, on one dataset (0.09 \emph{kWh} vs. 0.68 \emph{kWh}) without achieving a higher average and normalized \emph{nDCG@10} score.
One comparison that further illustrates the energy consumption and performance trade-off between deep learning and traditional algorithms is the comparison between the \emph{MacridVAE$^{RB}$} (deep learning) and the \emph{UserKNN$^{LK}$} (traditional) algorithms. 
While both algorithms achieve around the same normalized and averaged \emph{nDCG@10} performance (1.8 vs 1.7), \emph{MacridVAE$^{RB}$} consumes almost 60 times more energy (1.79 \emph{kWh} vs. 0.03 \emph{kWh}).

As widely known and highlighted by current research, hyperparameter optimization and randomness impacts the algorithm performance \cite{10.1145/3604915.3609488, WegmethVPB23}.
We acknowledge that tuning hyperparameters and repeating experiments could alter our results. 
However, this would likely increase the energy consumption disparity between deep learning and traditional algorithms. 
For instance, if we optimize the hyperparameters of \emph{MacridVAE$^{RB}$} and \emph{UserKNN$^{LK}$} through a grid search with 16 configurations and repeat the process with five different seeds, \emph{MacridVAE$^{RB}$} would consume 143.2 \emph{kWh}, while \emph{UserKNN$^{LK}$} would only use 2.4 \emph{kWh} (factor of 60).

The energy consumption difference between deep learning and traditional algorithms is not solely due to GPU usage.
While GPUs consume more energy than CPUs, implementation and complexity also play a role. 
For instance, \emph{ItemKNN$^{RB}$} on a GPU consumes no significantly more energy than CPU-based \emph{KNN} counterparts(\cref{fig:relation}).

\subsection{Carbon Footprint and Trends}

\subsubsection{The Carbon Footprint of Experiments at ACM Recsys 2023}\label{energy RecSys23}

Based on our experiments, we estimate that running all ACM RecSys 2023 full paper experiments emitted 886.9 metric tonnes of \emph{CO\textsubscript{2}} equivalents. Our \emph{CO\textsubscript{2}e} estimation is based on the number of submissions and the conversion factors from \emph{kWh} to \emph{gCO\textsubscript{2}e}.

The carbon footprint of all ACM RecSys 2023 full papers is closely tied to the number of submissions received. 
The conference accepted 47 full papers out of 269 submissions. 
Since the submissions involved running experiments, every submission added to the total carbon footprint of the ACM RecSys 2023 experiments. 
Consequently, we account for all 269 submissions in our analysis.

Our carbon footprint estimation is further based on the \emph{world average} conversion factor of 481 \emph{gCO\textsubscript{2}e} per \emph{kWh} \cite{ember2024carbon}.
We estimate that a full paper experimental pipeline consumes, on average, 6,854.4 \emph{kWh} (\cref{RQ1}).
With the conversion factor of 481 \emph{gCO\textsubscript{2}e} per \emph{kWh}, the carbon emissions of the 2023 ACM RecSys conference experiments in metric tonnes of \emph{CO\textsubscript{2}e} are calculated as follows:
\[ 6,854.4 \emph{kWh} \times 481 \emph{gCO\textsubscript{2}e} \times 269 \, (submissions) = 886.9 \, TCO\textsubscript{2}e \]
To illustrate, 886.9 \emph{TCO\textsubscript{2}e} is the equivalent of 384 passenger flights from New York (USA) to Melbourne (Australia) \cite{https://doi.org/10.1002/advs.202100707}. 
Or the amount of \emph{CO\textsubscript{2}e} that, on average, one tree sequesters in 80,600 years \cite{UKGov2020}.

\subsubsection{The Geographical Impact on the Carbon Footprint}
Variations in energy generation across different geographical locations affect carbon footprint by as much as 1200\% (45 vs. 535 \gcotwoe). 

Each location has its own \emph{kWh} to \gcotwoe conversion factor, which reflects the carbon intensity of its electricity generation.
The conversion factor is based on the energy sources of the respective geographical location. 
For instance, we ran our experiments in a geographical location that mainly utilizes renewable energy sources such as wind and hydropower \cite{ZHONG2021812}. 
Unlike this geographical location, some regions in Asia 
depend on coal \cite{pressburger2022comprehensive}. 
As a result, our conversion factor from \emph{kWh} to \emph{gCO\textsubscript{2}} is twelve times lower than that of the Asian region (45 vs. 535 \cite{ember2024carbon}.

If all experiments from ACM RecSys 2023 had been conducted in Sweden, the carbon emission estimation would have been reduced to 83 metric tonnes (90\% less than 886.9 \emph{TCO\textsubscript{2}e}).
In contrast, running the experiments in Asia, our estimation would have increased our carbon emission estimation by 99.6 metric tonnes of \emph{CO\textsubscript{2}e} (886.9 vs. 986.5 \emph{TCO\textsubscript{2}e})
Since no paper from ACM RecSys 2023 reported the data center or location where the experiments were conducted, we used the \emph{world average} conversion factor of 481 \emph{gCO\textsubscript{2}e} per \emph{kWh}\cite{ember2024carbon} to convert \emph{kWh} to \emph{gCO\textsubscript{2}e} to estimate the carbon emissions.

The amount of \emph{CO\textsubscript{2}} emitted by experiments is not solely determined by the location.
Some data centers operate mainly on renewable energy regardless of their location. 
For instance, \emph{Amazon} reports that their data centers used for \emph{AWS} cloud services predominantly utilize renewable sources\footnote{\url{https://sustainability.aboutamazon.com/products-services/the-cloud}}. 
Consequently, these specific data centers may have a conversion factor from \emph{kWh} to \emph{gCO\textsubscript{2}e} that differs from the general rate of their location.

\subsubsection{The Hardware Impact on the Carbon Footprint}
\begin{figure}
    \centering
    \includegraphics[width=1\linewidth]{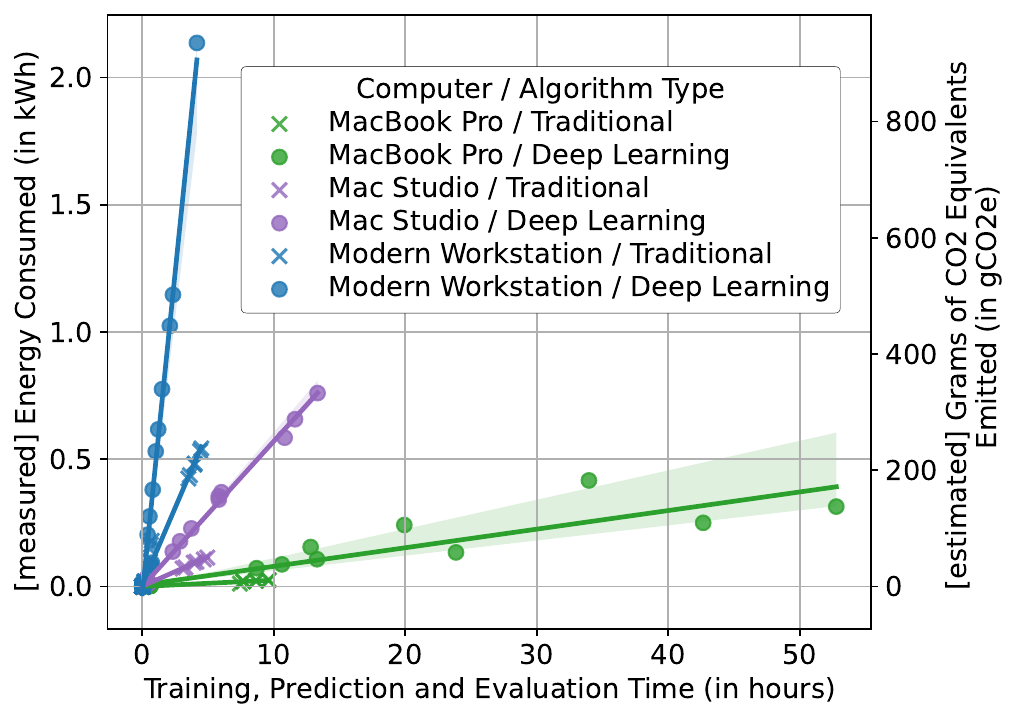}
    \caption{The relationship between energy consumption (in \emph{kWh}) and runtime, including training, prediction, and evaluation phases (in seconds). Each data point represents an algorithm applied to one of the twelve datasets. The linear functions illustrate the linear regression models for the respective groups of data points. The right-hand y-axis displays the corresponding \emph{gCO\textsubscript{2}e} emissions, calculated using the 2023 world average.}
    \Description[Test]{Test}
    \label{figure3}
\end{figure}

The \emph{CO\textsubscript{2}e} emissions of recommender systems experiments are influenced by more than geographical location, energy sources, algorithm types, and dataset characteristics. The type of hardware executing experiments can also affect the \emph{CO\textsubscript{2}e} emissions by a factor of up to ten. For example, the same experiments emit 14.4 \emph{gCO\textsubscript{2}e} when executed on an M1 MacBook Pro but 163.5 \emph{gCO\textsubscript{2}e} when executed on a modern workstation with an NVIDIA GPU.

Various hardware components, architectures, and cooling methods affect the energy consumption of recommender system experiments.  
Based on our experiments, a modern workstation with an NVIDIA GPU consumes, on average, five times more energy compared to an M1 Ultra Mac Studio (0.33 vs. 0.07 \emph{kWh}) and ten times more energy compared to the M1 MacBook Pro (0.33 vs. 0.03 \emph{kWh}).

Different hardware types do not only affect the energy consumption but also the running time of a recommender systems experiment. 
For instance, while a modern workstation uses, on average, ten times more energy than an M1 MacBook Pro, it completes the experiments, on average, in only one-third of the time (\cref{figure3}). 
Therefore, it is possible to save energy by running an experiment on an M1 MacBook Pro if you accept an increase in running time.

Overall, our results show a linear trend between energy consumption and runtime across various hardware types (\cref{figure3}). 
Although the slopes of the graphs vary, a consistent linear pattern is evident. 
This relationship suggests that a longer runtime is associated with increased energy consumption.

Even though a modern workstation consumes more energy, we run experiments on it because not all algorithms are compatible with Apple's ARM architecture.

\subsubsection{The Carbon Footprint of Recommender Systems 2013 vs. 2023}
\begin{figure}
    \centering
    \includegraphics[width=1\linewidth]{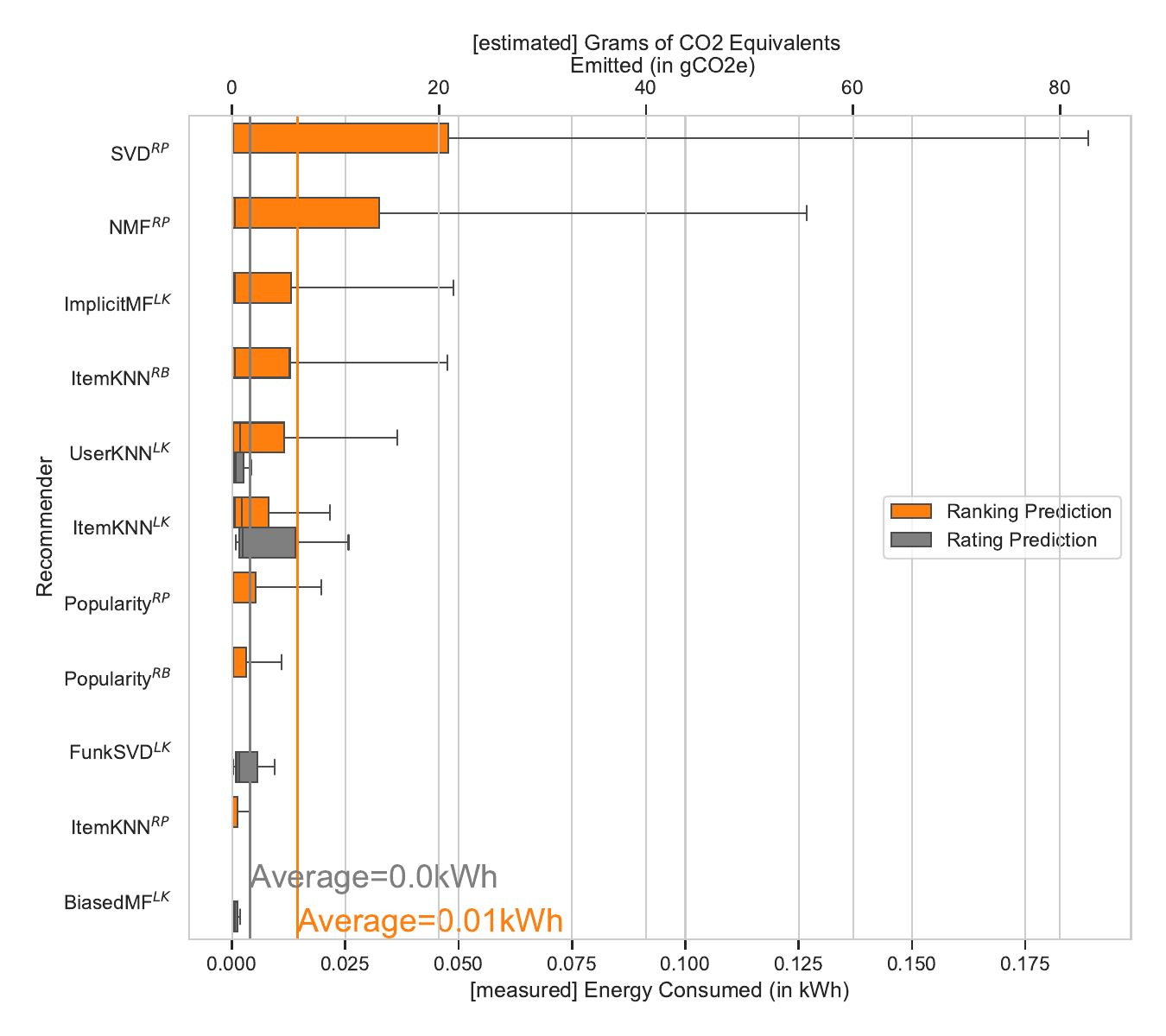}
    \caption{Average power consumption of traditional algorithms executed on 2013 hardware across seven datasets. The \emph{orange} vertical line indicates the average energy consumption in \emph{kWh} for ranking predictions and the \emph{orange} for rating prediction tasks. The upper x-axis shows the \emph{gCO\textsubscript{2}e} emissions, calculated using the 2023 world average. Not every algorithm is suited for rating- and ranking prediction tasks; therefore, not every algorithm displays two boxplots.}
    \Description[Test]{Test}
    \label{figure4}
\end{figure}

\begin{figure}
    \centering
    \includegraphics[width=1\linewidth]{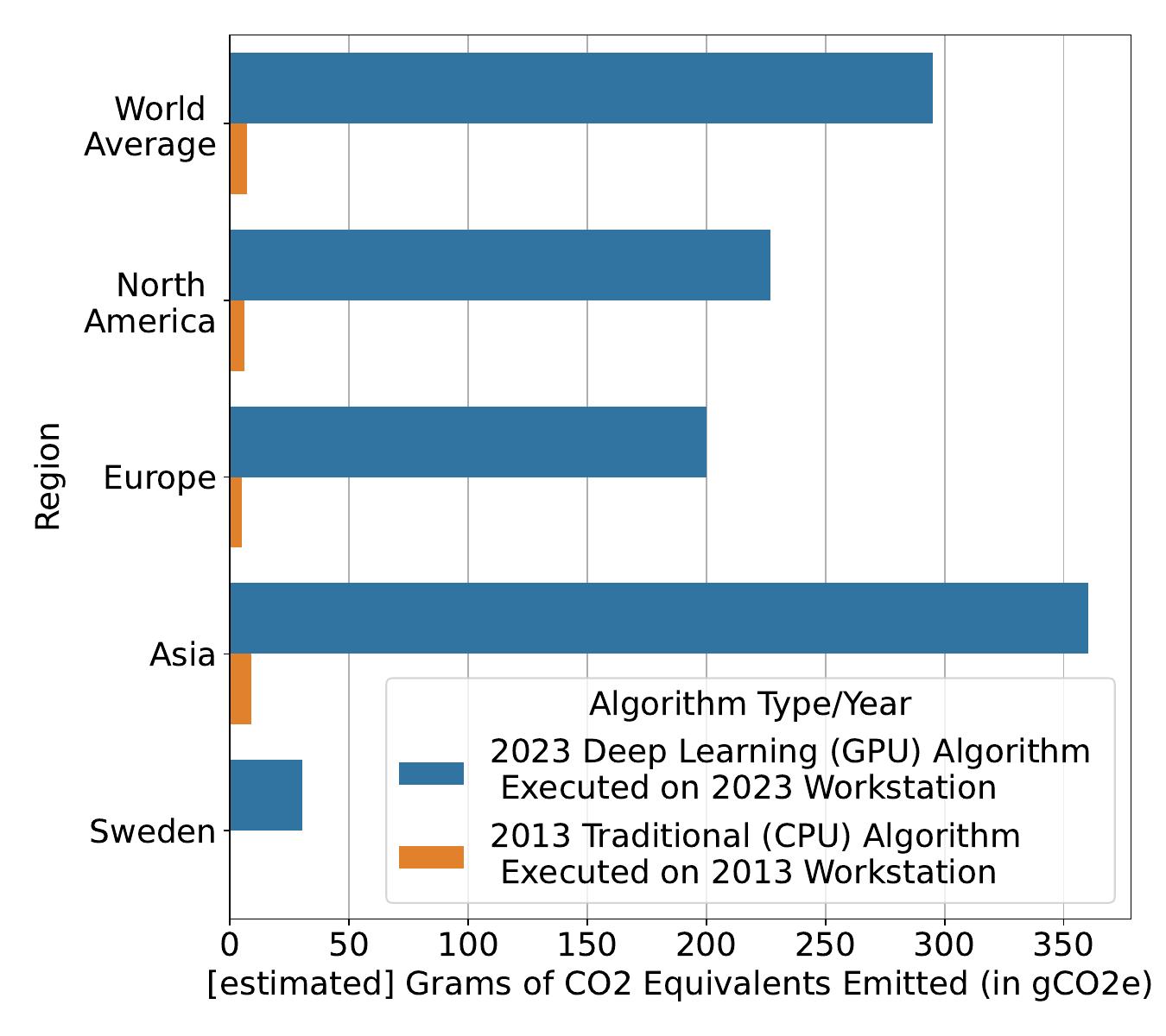}
    \caption[Test]{The regional variations in \emph{gCO\textsubscript{2}e} per recommender system algorithm type. The x-axis displays the \emph{gCO\textsubscript{2}e}, while the y-axis categorizes by region. Blue bars represent the emissions from a representative deep learning algorithm executed on the modern workstation hardware, and orange bars represent those from a traditional algorithm run on 2013 hardware. The \emph{gCO\textsubscript{2}e} are calculated using the respective annual conversion factors, reflecting changes in \emph{gCO\textsubscript{2}} per \emph{kWh} over the decade.}
    \Description[Test2]{}
    \label{figure5}
\end{figure}

The carbon footprint of recommender systems experiments in 2023 compared to 2013 has, on average, increased by a factor of 42 (7.09 \emph{gCO\textsubscript{2}e} in 2013 vs. 294,9 \emph{gCO\textsubscript{2}e} in 2023, \cref{figure5}, World Average). 

Simple ranking prediction algorithms used in 2013 recommender system experiments consumed, on average, five times less energy per run on one 2013 dataset using 2013 hardware (0.1 \emph{kWh}, see \cref{figure4}) compared to similar algorithms in 2023 executed on 2023 hardware (0.51 \emph{kWh}, see \cref{fig:avgpowcons}).
In 2013, no algorithm used, on average, more than 0.2 \emph{kWh} per run on one dataset using 2013 hardware (\cref{figure4}). 
In contrast, all deep learning algorithms in 2023 consume, on average, more than 0.2 \emph{kWh} per run on one dataset with 2023 hardware. 
The most energy-intensive algorithms in 2023 use more than six \emph{kWh} per run (\cref{fig:avgpowcons}).
While rating prediction was frequently used in 2013 recommender system experiments, the energy consumption for rating prediction is, due to prediction times, lower than for ranking prediction tasks (\cref{figure4}).

The shift towards more complex deep learning algorithms such as \emph{DGCF} or \emph{MacridVAE} and larger datasets like \emph{Yelp-2018}, compared to simpler algorithms like \emph{Popularity} or \emph{ItemKNN} and smaller datasets such as \emph{Hetrec-LastFM} used in 2013, can, in the worst case, increase energy consumption of more than 100,000 times.
For example, running the \emph{Popularity} algorithm on \emph{Hetrec-LastFM} and hardware from 2013 requires only 0.000049 \emph{kWh} while running \emph{DGCF} on dataset \emph{Yelp-2018} required 6.6 \emph{kWh} on 2023 hardware, i.e., a factor of around 134,000. 

The increased usage of clean energy sources and improved hardware efficiency in 2023, compared to 2013, does not compensate for the increase in carbon emissions due to the use of deep learning algorithms and larger datasets. 
A deep learning algorithm running on a 2023 dataset using 2023 hardware emits, on average, 42 times more \emph{CO\textsubscript{2}e} than a traditional algorithm on a 2013 dataset using 2013 hardware (7.09 \emph{gCO\textsubscript{2}e} in 2013 vs. 294,9 \emph{gCO\textsubscript{2}e} in 2023, \cref{figure5}, World Average). 
This estimation already includes the benefit of a better \emph{kWh} to \emph{gCO\textsubscript{2}e} conversion factors, improved by 62 \emph{gCO\textsubscript{2}e} due to the more frequent use of clean energy resources \cite{ember2024carbon}.

It is important to highlight that our analysis used hardware and algorithms from 2013 but not the exact software implementations used in the original papers. 
It was infeasible to retrieve old, unmaintained, non-centrally hosted software.
Consequently, many experiments likely used self-implemented algorithms. 
Compared to self-implemented algorithms, standardized software libraries could potentially change the efficiency, suggesting that the observed disparity in energy consumption between traditional and deep learning models might diverge.
\section{Discussion}
Our analysis highlights the significant environmental impact of full papers from the 2023 ACM Recommender Systems conference. 
We found that deep learning algorithms, when compared to traditional machine learning, consume substantially more energy without necessarily delivering better performance.
Additionally, the carbon footprint and environmental impact of recommender systems experiments have dramatically increased over the past decade.

We do not advocate abandoning deep learning algorithms but aim to raise awareness about the environmental impact of the trend toward deep learning-focused research. 
We encourage researchers and practitioners to document the experimental pipelines, computational overheads, hardware, and energy consumption in their publications. 
These details can help in understanding the environmental impact, highlight the energy demands, and reveal potential areas for energy efficiency improvements and reproducibility of recommender systems experiments.

Furthermore, we emphasize the importance of carefully selecting algorithms and datasets for recommender systems experiments. The environmental impact of deep learning algorithms can notably differ depending on the chosen algorithms and datasets. 
Researchers can minimize unnecessary computations by using efficient hardware or designing experimental recommender systems pipelines and thus reduce their environmental impact.
We also draw attention to the impact of hardware and geographic location on recommender systems experiments. 
If computers in different geographic locations are available, comparing the efficiency, hardware requirements, and energy source of the specific location can help reduce the environmental impact of running the experimental pipeline.
\section{Conclusions}
We reveal that the energy consumption of an average recommender systems research paper is approximately 6,854.4 \emph{kWh} (\textbf{RQ1}). 
Deep learning algorithms consume, on average, eight times more energy than traditional algorithms without achieving higher performance (\textbf{RQ2}). 
The carbon footprint of recommender systems experiments has increased significantly, with experiments from 2023 emitting approximately 42 times more CO\textsubscript{2}e when compared to experimental pipelines from 2013 (\textbf{RQ3}).

We want to raise awareness about the significant environmental impact of deep learning-focused research.
It is crucial that future publications include thorough documentation of the entire experimental pipeline, including computational overhead, hardware specifications, and energy consumption.
This transparency is not only essential for enhancing reproducibility but also for identifying potential areas for energy efficiency improvements.

Moreover, careful selection of algorithms and datasets is crucial, as the environmental impact of deep learning algorithms can vary significantly. 
Researchers can design more efficient experimental pipelines to minimize unnecessary computations and reduce environmental impact. 
Additionally, considering the impact of hardware and geographic location on experiments is vital. 
Comparing the efficiency, hardware requirements, and energy sources across different locations can further mitigate the environmental impact of experimental pipelines. 
We conducted all our experiments in Sweden, utilizing approximately 6,000 kWh of electricity, which corresponds to about 271.6 kilograms of CO\textsubscript{2}e. 
We planted 42 trees with One Tree Planted to offset our carbon emissions. 

Raising awareness about energy consumption and environmental impact can catalyze the development of more sustainable practices, benefiting the environment and recommender systems.

\begin{acks}
This work was in part supported by: \textbf{(1)} funding from the Ministry of Culture and Science of the German State of North Rhine-Westphalia, grant no. 311-8.03.03.02-149514, and \textbf{(2)} two ERASMUS+ Short-Term Doctoral Mobility Traineeships from the European Commission.
Furthermore, we are grateful for the support of Moritz Baumgart.
\end{acks}

\bibliographystyle{ACM-Reference-Format}
\bibliography{sample-base}

\end{document}